# How do digital threats change requirements for the software industry?

Veikko Halttunen

University of Jyväskylä, PO Box 35, FI-40014 University of Jyväskylä, Finland
`veikko.halttunen@jyu.fi`

**Abstract.** Digital systems are, by definition, the core of digital transformation. This has led many to think that the system being considered in digital transformation is solely software. I argue that this approach is a fatal mistake, and it has induced a great number of already realized problems and even a greater number of concerns about the future. These problems and concerns have become evident along with rising requirements for sustainability and responsibility. In this paper, I call for a better understanding of the digital society in its entirety. By digital society I mean the societal system that is affected by the digital systems and the ongoing societal transformations. When shifting the focus to the effects of digital systems on societies, we are forced to consider all the anticipated outcomes, both desirable and undesirable ones. Unfortunately, the mainstream research has ignored, to a large extent, the potential threats, and unwanted outcomes, of digitalization, which makes the efforts to change software businesses to be more sustainable, difficult to succeed. In my paper, I will provide an overall picture of current and future challenges of digital societies and discuss what these challenges mean to the software industry in future. The easiest way to start with, is to learn from earlier experiences, especially from the unsuccessful stories.

## 1   Introduction

Today, sustainability and responsibility are, or should be, among the leading trends of most industries. This also applies to the software industry. However, do we have a clear idea of what it means to be sustainable and responsible in software businesses? In this paper, I attempt to answer this question through a wide sociotechnical approach.

Digital systems have been considered as a crucial means to accelerate sustainability [1]. The promises, at least, have been grand. Instead, rare are the researchers and thinkers who really have tested the impacts of software industry on environment and societies. When going from a promise to another without resolving the existing problems and being prepared to new ones, the industry can hardly become sustainable and responsible.

There are two big ongoing changes that need to be considered together: the green transition, or more generally, the sustainability transformation(s) and the digital transformation. Both are fundamental societal processes that are linked with each other [2, 3]. In the context of digital systems several terms with different meanings are used for the transformation process [4]. *Digitization* is a technology-oriented term that refers to the change from analog information to digital form. *Digitalization* and *digital transformation* are terms with much broader meaning. Both refer to the manifold and wide changes digitization, i.e. implementation and use of digital systems, cause in societies. In [4], digital transformation is seen as the widest term being defined as "a fundamental change process, enabled by the innovative use of digital technologies accompanied by the strategic leverage of key resources and capabilities, aiming to radically improve an entity and redefine its value proposition for its stakeholders." An entity can be an organization, a business network, an industry, or society. As we look at the digital transformation on the societal level, the result of digital transformation is a digital society.

Green transition, which is a key topic in political discussions, is a process which primarily targets at shifting towards economies whose energy sources are other than fossil fuels. This is a crucial part of reduction of GHG-emissions and among the most acute actions that need to be taken. However, sustainability transformation includes many other goals that are associated with environment and societies [4]. These together with the above-mentioned impacts of digital transformation form a complex whole that must be approached in a systemic way.

Being sustainable/responsible means avoiding behavior that is unsustainable and irresponsible. By now, it is clear that discussion on sustainability is a result from the humans' unsustainable behavior. There is no other way than react, and learn from, earlier mistakes. In this paper, I follow this guideline and attempt to give some plain advice to the software industry. We already have a lot of information on features that may make a software product or service unsustainable/irresponsible. Leaving these features untouched and adding some other properties, which are considered as sustainable/responsible, means a trick that does not resolve the initial problem but just try to hide it.

My analysis is based on material collected for a new university course ('Introduction to sustainable and responsible digitalization') which was provided in June 2023. The course covered a large extent of problems and threats associated with digital systems. A great majority of the material was high-ranked research articles. For research-ethical reasons, I remark that the same material and examples generated from the material have been used for other research purposes[1] beside this paper. Since it was quite an effort to collect the material, I find it useful to share the results with scholars and practitioners.

In the following chapters, I first characterize well-known problems and threats related to digital systems or digitalization of societies. Next, I outline how this may affect the software industry as a primary actor of digitalization. Finally, I conclude my paper by providing a couple of normative ideas of how to continue.

## 2   Big picture of digital threats

In this chapter, I present an overall picture of problems digitalization has already caused. I also discuss some well-argued concerns and threats that may come true without a proactive approach towards digital societies. Awareness of threats of leaving the development to markets alone, is rising rapidly. This, in turn, delineates the frames for the software industry in future. Some of the new rules can be affected by the industry itself but some others must be taken as determined by the society. I consider this distinction in more detail in Chapter 3.

---

[1] Halttunen, V. (2023). Through a better understanding of 'undesirable futures' towards better digitalization https://ceur-ws.org/Vol-3598/paper15.pdf

Starting with the ecological issues, I go further by considering other areas that have been negatively affected by digitalization. These areas include digital divide, privacy concerns, misbehaviors of individuals, and problems related to health and well-being. I like to remind that many of these problems can be intertwined, which suggests that comprehensive research would be needed to analyze these interdependencies.

**Environmental impacts and concerns** have been a (or perhaps the) primary impetus for looking at digitalization through new lenses, see [5]. It has been argued that digitalization could be a pivotal factor in green transition [6]. However, outcomes so far have been modest, e.g. [7]. It seems that digital systems that, in theory, could facilitate green transition, have in practice caused increasing emissions through direct and indirect impacts [7]. The direct impacts come today largely from explosively increasing needs for data transfer. A great deal of data transfer capacity is required for amusement purposes, such as video streaming [8]. Blockchain validation processes are very energy consuming [9, 10], and new applications based on blockchain technology are planned and implemented with a rapid pace. The indirect negative impacts of digitalization on environment comes from boosting other industries that rely on energy consuming and/or polluting production and delivery of products/services [7]. One example of these is Shein, a Chinese ultrafast fashion retailer whose business is based on utilization of AI, social media and efficient production and logistic systems.

Lifecycles of the products in the ICT sector tend to be short [5]. IT devices requires rare raw materials that are often produced in an energy intensive way and causing environmental damages. The problem of short lifecycles is not, however, limited to the hardware business but applies to the software industry as well: all products and services should be designed to be maintainable and scalable.

**Digital divide** is a well-known problem caused by information and communication technologies. Digital divide means three things [11]. First, it is the gap between those who have proper ICT devices to access to the Internet services, and those who do not have. Second, it is the gap between those who possess appropriate skills and knowledge to use the devices, and those who don't. Finally, it is the gap between those whose environment make it possible to benefit the ICT based opportunities, and those whose environment don't enable this. Digital divide is a global problem, but it is also a problem of each society. like a recent from the UK shows [12]. Digital divide can be seen because of several reasons. Generally, it has been considered as societal issue. Nevertheless, the software industry may worsen the divide by accelerating the lifecycles of software products which may result in a technological race in which the poor have much less opportunities to succeed.

**Privacy concerns and data security threats** may be the biggest hindrance to digital transformation in future. There are several severe cases of privacy violation that may lead to resistance to public and private Internet-based services. One example is the Vastaamo case [13, 14]. Vastaamo was a Finnish psychotherapy center, whose patient files were hacked a few years ago, and the company and the victims extorted by threatening to publish the very sensitive data. Finally, as the extorter did not obtain what he wanted, he published patient data of thousands of patients, including the notes of patient sessions. What makes this case even worse, is that it happened to persons who were especially vulnerable due to their health problems. Concerning that low motivation to use digital services may lead to digital divide [15], it is necessary to ask whether cases such as Vastaamo improve to use digital services that deal with very sensitive information like that.

Privacy concerns are not limited to privacy violations, i.e. illegal use of private information. They also reflect the (usually) legal way of using customers data for commercial purposes. Zuboff, who call this surveillance capitalism, see big risks in this of digitalization [16]

**Cyberbullying, harassment and sexting** are rapidly increasing problems in digital environments [17-23]. Social media and Internet gaming are typical arenas where these undesirable and often illegal actions take place. Again, the situation is worst when the victims are unable to protect themselves, like students in schools, for example. Harassment, sexting and dating aggression are rather common phenomena among adolescents and adults who use digital devices and services like smartphones and applications of social media.

**Digital piracy** is an alternative term for infringement of copyrighted digital contents, such as music, videos, books, etc. [24] During the time of downloading music from the Internet, digital piracy was seen big problem by the music industry, e.g. [25]. Streaming-based services have improved the situation, but other problems have occurred (unfairness in distribution of income, for example). Digital piracy is, however, a persistent problem [24] that complicates the development of digital content services. For example, even when a streaming service is accepted by the customers, it may be considered unfair by the content providers [26].

**Health and well-being concerns related to digital systems** are manifold. They include both mental conditions (e.g. Internet gaming disorder, and other forms of addictive behavior) and physical conditions (e.g. obesity, neck pain, back pain, hearing problems etc. caused by computer game addiction). Addictive behavior is becoming a significant problem in digital societies [27,28]. Sometimes, a digital product can be both useful and harmful at the same time. For example, a videogame that may generate addictions, can be useful for learning purposes, for example [29]. Due to the conflicting effects, applications should be developed with care.

At organizational and interpersonal levels, technostress is currently a well-known problem identified by researchers [30]. Technostress worsens personal life and performance in work. A primary cause for technostress is the feeling that one is forced to be available around the clock.

## 3      Sustainable and responsible software industry

In respect of a specific digital system, sustainability and responsibility may be considered from a single aspect (or a few aspects) of the mentioned factors. Thus, some leeway has been granted for software businesses to develop their products. Considering digital systems as societal systems, the situation is different. All aspects of sustainability and responsibility must be taken into account at the same time. This implies that resolving one problem cannot cause or worsen some other problems.

Dealing with the problems of ICT-based societal systems, two instruments are available: (1) regulation by the society, and (2) improvements in business models and in technologies of the ICT sector. In this presentation, I do not regard the societal measures, and with regard to the ICT sector, I also exclude the measures that can be left to hardware providers. Thus, in the rest of the paper, I focus on what is required from the software industry in the future.

A systemic understanding of sustainability/responsibility issues is necessary for building a sustainable/responsible software industry. I have collected sustainability/responsibility problems relevant to software industry in Table 1. The collection of problems, and requirements induced from them, is not to be exhaustive but it rather aims to be a starting point for discussion on the industry's next steps towards sustainability and responsibility. Besides, it attempts to depict how large is the scope of sustainability/responsibility and how the different areas may interact with each other.

Table 1 describes the problems that are related to business models, products, product lifecycle and implementation methods. They are considered from different perspectives (environmental, social/societal, and individual). For example, the software *products* which are made plainly for amusement purposes, can be seen as an *environmental problem*. Many of these products are based on graphic processing, which is energy consuming even when used offline, but the problem is much worse when we talk about Internet games, for example.

**Table 1.** Sustainability and responsibility in the software industry – areas of concern, problems, and requirements (Problems: E - environmental, S - social/societal, I – individual)

| Area of concern | Problem(s) | Requirement(s) |
| --- | --- | --- |
| I. Business models | 1. Boosting unsustainability or irresponsibility in business ecosystems (E) <br> 2. Business models that are based on hooking or alluring etc. (I, S) | A) Software industry must consider all the effects of its business models on sustainability and responsibility of other businesses <br> B) The industry must avoid business models that are based on unethical ways of affect individuals or social groups |
| II. Products | 1. Products made for amusement purposes (E) <br> 2. Unsafe products (S, I) <br> 3. Addictive products (I, S) | A) Shift from amusement to necessities <br> B) New product innovations that have focus on security, human rights and well-being |

66

| | | |
|---|---|---|
| III. Lifecycle | 1. Consumption due to short lifecycles of products and services (E)<br>2. Short lifecycles worsen digital divide (S, I) | A) Software industry must target at longer lifecycles |
| IV. Methods | 1. Efficient to produce software but inefficient to run/maintain (E)<br>2. Emphasis on efficiency and productiveness instead of quality (e.g. data security) (S, I, E) | A) Software industry must prioritize programming languages (and other methods) that produce energy efficient code and support long product lifecycles through easiness to maintain.<br>B) New methods must improve data security and safety as a primary target. |

## 4 Conclusions

The software industry has been in a situation where its business models and products have not properly been challenged from perspectives of sustainability and responsibility. I am convinced that this will change quickly in the future. Hence, the industry would act wisely when it takes seriously all the aspects of sustainability/responsibility.

The growth of the industry has been to a large extent unsustainable/irresponsible. A good example of this is the gaming business. It has boosted IT use for amusement, which is typically energy consuming, providing few to no positive impacts on environment. It has also caused many kinds of individual and social problems, like addictive behavior, cyberbullying, and physical and mental health problems. This is not to say that gaming could not have any positive impacts. There are research findings that show positive effects of gaming on learning, for example. But even these effects must be evaluated with care against the possible risks.

I conclude my paper by the following message to the software industry. Be brave and ambitious to take seriously the sustainability/responsibility challenges and be ready to reform from inside. It implies that the research agenda must be updated by including critical research issues on the list. This is the only way to align digital transformation with the necessary environmental and societal changes.